# On sparse sensing and sparse sampling of coded signals at sub-Landau rates

Michael Peleg[(1,2)] and Shlomo Shamai[(1)]

[(1)]Technion-Israel Institute of Technology, [(2)]Rafael Ltd.

ABSTRACT

Advances of information-theoretic understanding of sparse sampling of continuous uncoded signals at sampling rates exceeding the Landau rate were reported in recent works. This work examines sparse sampling of coded signals at sub-Landau sampling rates. It is shown that with coded signals the Landau condition may be relaxed and the sampling rate required for signal reconstruction and for support detection can be lower than the effective bandwidth. Equivalently, the number of measurements in the corresponding sparse sensing problem can be smaller than the support size. Tight bounds on information rates and on signal and support detection performance are derived for the Gaussian sparsely sampled channel and for the frequency-sparse channel using the context of state dependent channels. Support detection results are verified by a simulation. When the system is high-dimensional the required SNR is shown to be finite but high and rising with decreasing sampling rate, in some practical applications it can be lowered by reducing the a-priori uncertainty about the support e.g. by concentrating the frequency support into a finite number of subbands.

*__Correspondence__
Michael Peleg E-mail: peleg.michael@gmail.com

## 1 INTRODUCTION

Sparse sampling and sparse sensing comprise an active and broad research field applied also to radio communications. Extensive research has been reported treating the case of random Gaussian signals highly relevant to wireless communications, e.g. the recent [1], [2], [3] and references within. In this work which is an extension of [4] we examine the impact of introducing coding into the sparse signal. The results may be applicable to reducing further the sampling rates when receiving sparse and coded communication signals and also to exposing the dependence of support identification on the nature of the signals in other sparse sensing scenarios.

We examine a setting in which a frequenc[1]y- sparse signal occupying only some frequencies and subbands out of the available bandwidth $W$ is transmitted over an Additive White Gaussian Noise (AWGN) channel. The receiver desires to recover both the signal frequency support and the signal itself. If the signal is uncoded it is well established that to acquire the noisy signal the Landau condition [5] requiring the sampling rate to exceed the occupied non-contiguous bandwidth must be met even with known frequency support. The task of jointly detecting the frequency support of the signal and to acquire the signal itself requires somewhat higher sampling rates. The works [1], [6], and [7] show that at high enough Signal to Noise Ratio (*SNR*) a sampling rate only slightly higher than the Landau rate is sufficient with high probability when using high performance detection algorithms while somewhat higher sampling rates are required at moderate SNR or when the algorithm complexity is constrained [2], [8], [9]. It is well known, .e.g [6], that the sparse sampling problem can be treated as a sparse sensing as explained also in the next section. Then the frequency support is treated as the support of the sparse signal and the sampling rate is treated as the rate of measurements. Recent works [1], [2] and [3] examined the fundamental limits on rates of measurements of sparse sensing in the asymptotic high-dimensional case of many measurements; we summarize

---

[1] Submitted to the ETT, *Transactions on Emerging Telecommunications Technologies*

here the results relevant to this work. The paper [1] analyzes optimal decoders of sparse continuous signals defining the concept of continuity rigorously using measures such as the Renyi information dimension. For large number of measurements and with optimal decoders, [1] showed that the number of measurements required and sufficient both to achieve errorless reconstruction in the noiseless case and to achieve a finite noise sensitivity is equal to the support size and that this holds also with random Gaussian sensing matrices. This corresponds to sampling at the Landau rate [5]. Interestingly an additional non-sparse discrete (e.g. with a finite cardinality alphabet) component [1, eq. (3)] of the signal is detectable together with the sparse continuous signal, in some cases without increasing the required number of measurements. Such a discrete component can model a coded signal as examined in this paper. The works [2] and [10] analyze support recovery when a small error rate is permitted enabling support recovery with the linear sparsity measurement rate defined in [2, eq. (8)]. They provide results for a variety of signal models and SNRs and the references within treat support recovery under different scenarios and requirements. The results most relevant to this work are that under most scenarios the number of measurements required for small error rate approaches the support size at asymptotically high SNR and exceed it significantly at general SNR. Still the detection of the support with additional discrete signal is feasible even when the joint support size of the continuous and discrete signals exceed the number of measurements, [2, eq.(57), (59)]. In [3] the information theoretical limits with Gaussian inputs are studied. Mutual information and SNR requirements in the high-dimensional limit are derived for a class of sensing matrices. It shows [3, theorem 6] that the Landau condition is necessary with continuous uncoded Gaussian input even in the noiseless case.

Thus the works cited above focus on continuous signals sampled faster than the Landau rate and predict but not analyze in detail a significant reduction of the required sampling rate when the uncertainty about the signal is dramatically reduced by coding. This paper focuses on coded sparse signals sampled at rates below the Landau rate and exploits coding for reliable reconstruction of the signal and of its support. We demonstrate that with coded signals the joint detection of the support and of the signal can be achieved with the number of measurements smaller than the support size corresponding to a sampling rate below the Landau rate. We examine the information-theoretic tradeoffs between SNR, coderate, signal sparsity and the required sampling rate. Our results hold for general, not necessary asymptotically high, number of dimensions. The receiver cannot be decomposed into sparse sensing followed by decoding as in the uncoded case, rather a joint support and signal recovery is required to exploit the code in the support recovery. This is since the constraints on the signals introduced by the coding are exploited to overcome the lower sampling rate. In the asymptotic high dimensional case the required SNR becomes very high similarly to the continuous case e.g. [2], [3] due to the high entropy of the support but it can be lowered by reducing the a-priori uncertainty of the frequency support which is common in radio communication systems when bandwidth is assigned in a finite set of subbands.

The paper outline: in section 2 we present the system model and derive bounds on information rates, section 3 demonstrates reliable detection achievability with the support known to the transmitter, section 4 analyzes and simulates a system in which the support is not revealed to the transmitter and section 5 examines the asymptotic high-dimensional case of many measurements and its relevance to communication systems operating at moderate SNR. Section 6 concludes the paper. All logarithms are base 2.

## 2 CHANNEL MODEL AND PRELIMINARY OBSERVATIONS

### 2.1 The model

We wish to sense a sparse vector **x** with $n$ elements $x_i$, out of which $q$ are non-zero, from a measurement **y** which is a length-$p$ vector produced by the sensing

$$\mathbf{y} = \mathbf{A}\mathbf{x} + \mathbf{z} \qquad (1)$$

The sensing matrix **A** is of size $p,n$ and its elements are drawn identically and independently distributed (i.i.d.) for each use of the channel (1) from a complex Gaussian distribution with a unit variance. The matrix **A** is known to the receiver but not to the transmitter. The non-zero elements of the signal vector **x** are complex with a unit mean square value (average power). The noise **z** comprises i.i.d. complex Gaussian elements of variance $q/SNR$. Clearly the signal to noise ratio of each element of **y** is *SNR*. The $q$ positions of the non-zero elements of **x** are denoted the support $b$ of **x** and $p$ is the number of measurements where $n$, $p$ and $q$ are general and finite. The support $b$ is information desired by the receiver and it cannot be influenced by the transmitter. We shall examine the cases of support known and unknown to the transmitter.

By modifying **A**, the model (1) is directly applicable to sparse sampling of finite duration signals if the support *b* represents the frequencies used, the elements of **x** are frequency domain symbols transmitted over those frequencies, the elements of **y** are the time domain samples at the receiver and **A** comprises the rows of the Inverse Fourier Transform (IFT) matrix corresponding to those samples. Then the Landau condition [5] stating that the sampling rate must not be smaller than the frequency support corresponds directly *to p ≥q*. If the sampling times are a subset of the regular Nyquist rate sampling times then the rows of **A** are orthogonal rows of the IDFT (Inverse Discrete Fourier Transform) matrix.

In the following we shall analyze systems with a Gaussian **A** corresponding to the model used widely for sparse sensing. We shall analyze performance also with **A** comprising *p* rows of the discrete Fourier transform selected randomly and uniformly at each use of the channel (1) corresponding to the sparse sampling problem.

Let's describe **x** by means of it support *b* and by its other information content *c* referred to also as the signal using the function *f*:

$$\mathbf{x} = f(b, c) \quad (2)$$

We examine two possible constructions of **x**.

*Construction C1*:

Define $c = \mathbf{x}^0$ as a length-*q* vector comprising the nonzero elements of **x**. Define *b* as an ordered set of *q* indices of the nonzero elements in **x**. Initialize **x** as a vector of zeros and then assign the elements of $\mathbf{x}^0$ to the positions in **x** indexed by the elements of *b*. Define $\mathbf{A}^0$ as the *p,q* submatrix comprising the corresponding columns of **A**. Then (1) can be re-written as

$$\mathbf{y} = \mathbf{A}^0 \mathbf{x}^0 + \mathbf{z} \quad (3)$$

*Construction C2:*

Define *c*=**c** as a length-*n* vector of complex elements according to the desired Probability Density Function (PDF) of the elements of **x**. Define *b* as a length *n* vector of *n-q* zeros and of *q* ones designating the positions of the non-zero elements of **x**. Then construct **x** by element-wise multiplication of **c** and *b* as done in [3].

We are interested in the information rates *I*(**y**;**x**|**A**), *I*(**y**;*b*|**A**) and *I*(**y**;*c*|**A**) under various PDFs of **x**. Using (1), (2) and the fact that $P(\mathbf{y}|\mathbf{x}, \mathbf{A}) = P(\mathbf{y}|b, c, \mathbf{A})$ where *P* denotes probability yields:

$$I(\mathbf{x}; \mathbf{y} | \mathbf{A}) = I(b, c; \mathbf{y} | \mathbf{A})$$

Then by the mutual information chain rule similarly to [3]:

$$I(\mathbf{x}; \mathbf{y} | \mathbf{A}) = I(c; \mathbf{y} | \mathbf{A}) + I(b; \mathbf{y} | c, \mathbf{A}) \quad (4)$$

A central issue is the information retrievable about **x**, its partitioning between the support *b* and the content *c* and the influence of the sparsity parameters *q* and *p*. We denote the information content (the entropy) of *b* and the information rate carried by *c* as $R_b$ and $R_c$ respectively.

With Gaussian i.i.d. $\mathbf{x}^0$ it is well known that for *q< p* the vector **x** comprising the support *b* and the signal *c* can be recovered from **y** with reasonable reliability under some conditions, e.g. [1], [11]. For *q>p* it is known not to be the case since it violates the Landau condition [5]. In this paper we examine under which conditions is the support and signal recovery possible in the region *q>p*. Particularly we shall examine the case where $\mathbf{x}^0$ is chosen from a finite alphabet such as done in coding. The mutual information terms in (4) are interesting by their own merit, to further demonstrate the feasibility of reliable decoding even with a small *n* we shall use codes and allow the use of the channel (1) for *N* times as required by the code length. We show in section 3 below that when the support is known at the transmitter it can be recovered reliably at the receiver using the state amplification technique presented in [12] despite it not being under control of the transmitter. This enables operation at SNR lower than those of [2] derived without the coding.

### 2.2 Initial analysis

The information rate *I*(**x**;**y**|**A**) in (4) is central to our problem; we shall denote its maximal value under each scenario by $I_M$.

$$I_M = \max_{P(\mathbf{x})} I(\mathbf{x}; \mathbf{y} | \mathbf{A})$$

Clearly, to decode *b* and *c* reliably at the receiver we need

$$R_b + R_c \leq I_M \quad (5)$$

We shall assume for ease of presentation that the support $b$ is distributed uniformly over all possible $b$'s constrained only by $q$ and $n$. The entropy of $b$ is then

$$H(b) = \log \binom{n}{q} = \log \frac{n!}{(n-q)!q!} \quad (6)$$

As defined above $R_b=H(b)$. Thus if the support and the signal are to be recovered, $R_c$ is limited to

$$R_c \leq I_M - \log \frac{n!}{(n-q)!q!} \quad (7)$$

Next we shall derive lower and upper bounds on $I_M$. The mutual information over the channel (1) with $\mathbf{A}$ known to the receiver is

$$I(\mathbf{x};\mathbf{y} \mid \mathbf{A}) = H(\mathbf{y} \mid \mathbf{A}) - H(\mathbf{y} \mid \mathbf{x}, \mathbf{A}) \quad (8)$$

The term $H(\mathbf{y}|\mathbf{x},\mathbf{A})$ in (8) is just the entropy of the noise independent of the transmitted signal as in the MIMO case with $p$ receive antennas analyzed in [13]. Thus the transmitter should maximize only $H(\mathbf{y}|\mathbf{A})$. Since $b$ is not influenced by the transmitter the conditional PDF of $\mathbf{y}$ is:

$$P(\mathbf{y} \mid \mathbf{A}) = \sum_b P(b) \sum_{\mathbf{x}^0} P(\mathbf{y} \mid b, \mathbf{x}^0, \mathbf{A}) P(\mathbf{x}^0 \mid b) \quad (9)$$

where $P$ denotes probability. The PDF $P(\mathbf{x}^0|b)$ has to be optimized for maximal $H(\mathbf{y}|\mathbf{A})$. With *Construction C2, c* must replace $\mathbf{x}^0$ in (9).

First we shall establish that for the Gaussian $\mathbf{A}$ and with *Construction C1*, $H(\mathbf{y}|\mathbf{A})$ cannot be increased by transmitting a signal $\mathbf{x}^0$ dependent on $b$. Note that the columns of $\mathbf{A}$ as defined above are i.i.d. Suppose the transmitter maximizes $H(\mathbf{y}|\mathbf{A})$ employing distribution denoted by $P^T(\mathbf{x}^0|b)$. Different strategies $P(\mathbf{x}^0|b)$ will impose different $P(\mathbf{y}|\mathbf{A})$ in (9). Note that $H(\mathbf{y}|\mathbf{A})$ is invariant to any permutation of $b$ in $P^T(\mathbf{x}^0|b)$ because the columns of $\mathbf{A}$ are i.i.d. and the support only assigns elements of $\mathbf{x}^0$ to columns of $\mathbf{A}$. Let us consider the set of all such strategies denoted $P^P(\mathbf{x}^0|b)$ derived from the actual $P^T(\mathbf{x}^0|b)$ by all the possible permutations on $b$, each achieving the same $H(\mathbf{y}|\mathbf{A})$. Due to the convexity of entropy, $H(\mathbf{y}|\mathbf{A})$ is a convex function of $P(\mathbf{y}|\mathbf{A})$ in (9). Thus $H(\mathbf{y}|\mathbf{A})$ will not decrease if the transmitter uses instead of $P^T(\mathbf{x}^0|b)$ a linear equally weighted combination of all the permuted strategies $P^P(\mathbf{x}^0|b)$. The equally weighted combination yields $P(\mathbf{x}^0|b)$ independent on $b$. Thus some transmit signal $\mathbf{x}^0$ with PDF independent of $b$ achieves $I_M$. This independence between $\mathbf{x}^0$ and $b$ holds for each single use of (1) and is compatible with the statistical dependence of sequences of $\mathbf{x}^0$ and $b$ introduced by coding in section 3 below. The same conclusion holds for *Construction C2* as is seen by replacing $\mathbf{x}^0$ by $c$ and following the derivation verbatim. Using $c$ independent of the support $b$ is found effective in section 2.3 below also with the Fourier matrix.

A lower bound on the achievable $H(\mathbf{y}|\mathbf{A})$ can be obtained by comparing to MIMO communication with $p,q$ channel matrix unknown to the transmitter and known to the receiver which maximizes $H(\mathbf{y}|\mathbf{A})$ by Gaussian i.i.d. signals, see [13]. Let a Gaussian i.i.d. $\mathbf{x}^0$ be transmitted over our channel. Our received $\mathbf{y}$, as formulated in (3), has a PDF different from the MIMO case in [13] since the unknown support changes the submatrix $\mathbf{A}^0$ of the known matrix $\mathbf{A}$. This difference is eliminated if $b$ is known at the receiver, in other words our $H(\mathbf{y}|\mathbf{A},b)$ is the same as $H(\mathbf{y}|\mathbf{A}^0)$ in the MIMO case [13]. This is a lower bound on $I_M$ because $H(\mathbf{y}|\mathbf{A}) \geq H(\mathbf{y}|\mathbf{A},b)$. Thus $I_M$ is lower bounded by the results on $I(\mathbf{x};\mathbf{y}|\mathbf{A})$ in the $p,q$ MIMO channel presented in [13]

$$I_M \geq I_{MIMO} \quad (10)$$

where $I_{MIMO}$ is the Average Mutual Information (AMI) of the MIMO channel with $q$ inputs and $p$ outputs with the channel known to the receiver but not to the transmitter analyzed in [13, theorem 2] and plotted for $q=6, p=3$ in figure 1.

The maximal $I_M$ is expected to exceed its lower bound somewhat since the unknown support $b$ is expected to increase $H(\mathbf{y}|\mathbf{A})$ relative to the MIMO case, so an upper bound on $I_M$ is of interest. The upper bound can be derived by releasing all constraints on $\mathbf{y}$ except its power. The maximal $H(\mathbf{y})$ of any vector $\mathbf{y}$ with our power constraint of $E(|\mathbf{y}|^2) = pq(1+1/SNR)$ is achieved by a vector with i.i.d. Gaussian elements leading to an upper bound on $I_M$ denoted here as $I_{MUP}$:

$$I_M \leq I_{MUP} = H(\mathbf{y} \mid \mathbf{A}) - H(\mathbf{z}) = p\log(1+SNR) \quad (11)$$

This upper bound coincides with the limit on $I_{MIMO}$ in [5, eq.(6)] if $p$ is fixed and $q$ is raised to infinity. In this work we have always $q>p$, so the gap between the upper bound and between the lower bound $I_{MIMO}$ is small. The bounds are plotted in figure 1 for $q=6, p=3$. Tighter bounds and even exact expressions are available in the asymptotic case of large $p$ and $q$, see [3], [14], [15] and references within discussed in section 5 below.

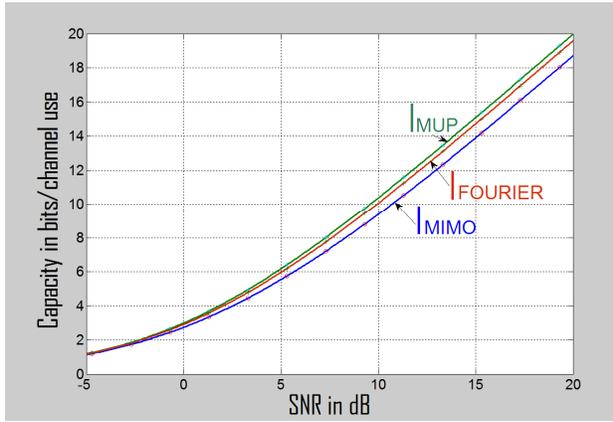

Figure 1. Lower and upper bounds on $I_M$ with $q=6$, $p=3$ with Gaussian matrix **A** and the lower bound with a Fourier matrix **A** for n=10.

## 2.3 Application to sparse sampling

The Gaussian matrix **A** applies to the most researched sparse *sensing* problem. We showed above that to treat the sparse *sampling* problem, **A** will comprise orthonormal rows selected from the IDFT transform matrix. The information rates turn out nearly identical in most cases. The upper bound (11) applies by its derivation which does not depend on the PDF of **A** but only on its second moments which are unity as in the Gaussian case. The lower bound can be found as proved above by conditioning on *b*:

$$I(\mathbf{x};\mathbf{y} \mid \mathbf{A}) \geq I(\mathbf{x};\mathbf{y} \mid \mathbf{A},b) \qquad (12)$$

which is again the capacity of a MIMO channel with $p,q$ matrix unknown to the transmitter and known to the receiver. Also, as with a Gaussian **A**, for $p \ll q$ the lower bound will reach the upper bound because the elements of **y** will become Gaussian i.i.d.

Since the capacity of a MIMO channel with a channel matrix **A** being a submatrix of the Fourier transform matrix is not available, we evaluated it numerically for the case p=3, q=6, n=10, selecting randomly the rows and columns of the submatrix while using the capacity formula [13]

$$I = E\log(|\mathbf{I}_p + \frac{SNR}{q} * \mathbf{A}\mathbf{A}^H|) \qquad (13)$$

Where $E$ stands for expectation, **I** is the identity matrix and |**G**| denotes the determinant of **G**. The lower bound using the Fourier submatrix is closer to the common upper bound (11) than when using the Gaussian matrix for the matrix dimension examined in Fig. 1, see the line denoted $I_{Fourier}$. Thus our conclusions about sparse sensing apply well to the sparse sampling too and the use of signal *c* independent of the support *b* nearly achieves the upper bound on mutual information as it did in the Gaussian matrix case.

## 2.4 The channel state perspective

The problem treated here is within the framework of channels dependent on a state. Specifically the support *b* which is not under the control of the transmitter can be treated as a channel state while the content *c* is treated as the information to be transferred over the channel.

The case with channel state known at the transmitter covering our case of support revealed to the transmitter enjoys a rich theoretical background starting with Shannon [16] and surveyed recently in [17]. The work most relevant to our problem is [12] which attempts to detect both *c* and *b*, the later known to the transmitter and derives the capacity region of the rates $R_c$ and $R_b$ using technique similar to Gel'fand-Pinsker [18]. The case with the side information known only casually is analyzed also in [19]. The result of [12] relevant to this work is mainly the capacity region [12, Theorem 1] defined in our notation by (5) above and by an additional constraint on $R_c$ identical to the Gel'fand-Pinsker capacity expressed in our notation by

$$R_c \leq I(U;\mathbf{y}) - I(U;b) \qquad (14)$$

However since in our case the support *b* is demanded to be perfectly detected, (14) collapses into (5) if $U=(b,c)$ is selected and substituted into (14):

$$R_c \leq I(b,c;\mathbf{y}) - I(b,c;b)$$
$$R_c \leq I_M - R_b$$

which yields (5), so the constraint (14) does not change the capacity region in our case. Thus in our case (5) is the capacity region and it follows from [12] that it is achievable. Since we always desire perfect reconstruction of *b* we can use a somewhat simpler achievability scheme which we present in section 3. This proves the following:

*Theorem 1*: With the support known non-casually at the transmitter a reliable detection of the signal and the support is possible and the capacity region of $R_b$, $R_c$ is given by (5) and (6) with $I_M$ lower and upper bounded by (10) and (11) respectively.

The case with side information not known at the transmitter covering our case of support unknown to the transmitter is analyzed in [20] which derives the capacity-distortion region. It is clear that in this case the channel state (support in our case) cannot be detected errorlessly and the use of distortion measure is in place. Theorem 1 in [20] presents the capacity-distortion region and demonstrates that the need to estimate the channel state at the receiver can be translated into a constraint on the PDF of the transmitted symbols.

## 3 ACHIEVIBILITY SCHEME WITH SUPPORT KNOWN NON-CAUSUALLY TO THE TRANSMITTER

The lower bound on the information rate $R_c$ with joint reliable detection of the signal $c$ and of the support $b$ given by (7) and (10) is achieved by a simple and straightforward communication scheme utilizing the classic random coding method. See figure 2.

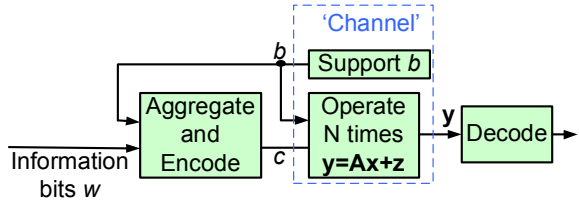

Figure 2. Achievability scheme with support known at the transmitter. The support is denoted by $b$, the transmitted signal by $c$.

The information word $w$ out of $2^{R_c N}$ possible words is encoded over $N$ uses of the channel (1). As defined above, **x** is a function of the pair $(b,c)$. Denote the series of $b$, $c$, **x**, **y** as $B$, $C$, **X**, **Y** respectively. The encoding function will be then $(B,w) \rightarrow$ **X**. It is constructed as a random like code with rate approaching the lower bound in *theorem 1* above. Thus the rates are selected as

$$R_b = H(b) \quad (15)$$

$$R_c = I_{MIMO} - H(b) \quad (16)$$

In practice the codebook would be created by drawing $C$ from an i.i.d. complex Gaussian distribution for each $(B,w)$ pair while $B$ was already "drawn" by the channel, yielding a codebook of $(B, C)$ pairs. For conformance to the standard coding argument a slightly modified codebook is required as follows. The whole codebook of $2^{NR}$ codewords $(B,C)$ is drawn at random where $R=R_b+R_c$. It is easy to show from (15) and from the number of codewords that each possible support sequence $B$ appears in the codebook with high probability at least $2^{R_c N}$ times, thus for any actual support $B_i$ a different codeword $(B_i,C)$ can be matched to each distinct $w$. Thus at transmission the $B$ in the codeword can be selected identical to the actual $B$ of the channel. This creates the same codebook as the standard random coding procedure regardless of drawing $B$ by the channel. Then the information rate (16) can be achieved while also detecting the support reliably by the standard random coding arguments: the mutual information over the channel is at least $R_b+R_c$ because the transmitted signal is drawn from the same PDF as in the section on lower bounds above, the channel (1) is memoryless and we are using a random-like code encoding $B$ and $C$ jointly into **X**.

Another construction of a coding scheme would be applying the technique of [12] as described in section 2.4 above. If the auxiliary variable in [12] is selected as $U=(b,c)$ then following the achievability proof of [12, Theorem 1] yields the same communication scheme as described above.

We used signals drawn from the Gaussian distribution because they achieve the lower bounds on information rates derived above and because those lower bounds are near the upper bounds. There is a possibility that there are better distributions. Indeed a work on limited duty cycle communication [21] showed that if the support is sparse (low duty cycle in [21] corresponding to $q<n$ here) and if the support is considered a part of the information, then discrete distribution outperforms the Gaussian one and achieves capacity. However the results [21] are not directly applicable here since [21] is not sparse sensing but rather uses the full sampling rate and the support $b$ there is under the transmitter control.

The results of this section are valid also for **A** being a submatrix of the Fourier matrix with the slightly different $I_{MIMO}$ calculated by (13) and presented in the high-dimensional case in section 5 below.

Following (16), the *SNR* required for reliable operation is a monotonically increasing function of $R_c$. For vanishing $R_c$ the *SNR* must provide $I_{MIMO} \geq H(b)$ which is, for the example $n=10$, $q=6$, about 7.96 dB using (6) and figure 1.

There are two possible extensions:

If the PDF of the support is not uniform as presented but rather follows some other distribution set by the channel then all the above derivation holds except for the support information rate given by (6) which must be modified accordingly. This will be used in the end of section 5 below.

If the support would be under the control of the transmitter the system would still operate and achieve the same rates since the ability to control the transmitter cannot reduce the reliable communication rate and since the upper and lower bounds on the mutual information over the channel, (10), (11) remain valid.

## 4 SUPPORT AND CHANNEL NOT KNOWN TO THE TRANSMITTER

### 4.1 The system description

The system structure is as in figure 2 while deleting the $b$ input to the encoding function. This yields a standard communication setting over a memoryless channel and standard random coding is used with no need for the slight modification used in the previous section.

In this case the support corresponds exactly to the state of the classical state dependent channel with state unknown to transmitter and receiver. The support cannot be identified with arbitrary high probability with noise present, e.g. [20]. Still the same rates as in (15), (16) are desired. An intuitive strategy can be transmitting information over the signal $c$ at the maximal rate which permits reliable decoding, so the receiver can decode the signal $c$ reliably and then utilize it for the detection of the support $b$. To do so the transmitter encodes over $N$ channel uses in a manner similar to the previous section and with the same information rate $R_c$, except that the encoded information now is $C$ instead of $(B,C)$. The same Gaussian PDF of $\mathbf{x}$ is used. The rate of the information $R_c$ to be reliably decoded is clearly limited still by (7) if the support is to be recovered with low error rate. Also the rate $R_c = I(c;\mathbf{y}|\mathbf{A})$ is achievable by standard random coding arguments without the need of the extension used in the previous section.

*Construction 1* as defined in section II above is applicable here but then the combining of the signal $c$ and the support $b$ in (2) is performed by the channel. Here *Construction C2* is a more intuitive alternative where the transmitter oblivious of the support produces a non-sparse $c=\mathbf{c}$ and lets the channel choose which components are actually transmitted. With both the constructions $c$ will be recovered reliably due to the coding as shown next. Let us derive a lower bound on the achievable $R_c = I(c;\mathbf{y}|\mathbf{A})$:

$$I(c;\mathbf{y}|\mathbf{A}) = H(\mathbf{y}|\mathbf{A}) - H(\mathbf{y}|\mathbf{A},c) \qquad (17)$$

If we condition both the entropies in the last equation also on $b$, both will decrease by up to $H(b)=R_b$ so

$$R_c \geq I(c;\mathbf{y}|\mathbf{A},b) - R_b$$

The mutual information in the last equation, if the transmitter uses i.i.d. Gaussian signals, is $I_{MIMO}$ thus we can always achieve

$$R_c = I_{MIMO} - R_b \qquad (18)$$

as in (16). Since $b$ is not coded it will not be recovered errorlessly and its reliability of detection is of interest. It is treated next.

### 4.2 Support detection performance

The receiver first decodes $c$ reliably using all the $N$ received vectors $\mathbf{y}$. When $c$ for particular channel use is known then the sufficient statistics to decode $b$ from is $\mathbf{y}$, $c$ and $\mathbf{A}$ of this channel use only. Although a practical receiver would produce an estimate of the support jointly with recovering $c$, the support detection performance can be analyzed using the reliably decodable $c$ at the support detector input.

The received signal (3) is a product of a random Gaussian matrix $\mathbf{A}^0$ with a random Gaussian vector $\mathbf{x}^0$. In our case the vector is the detected signal already known to the receiver and the matrix is selected by the support $b$ from the columns of the Gaussian matrix $\mathbf{A}$ which is also known to the receiver. Given $\mathbf{A}$ and $\mathbf{x}^0$ and without the additive noise, $\mathbf{y}$ is determined solely by $b$.

The error probability can be bounded and estimated numerically for small number of measurements as follows: with known $\mathbf{A}$ and $c$ the vector $\mathbf{A}\mathbf{x}=\mathbf{A}^0\mathbf{x}^0$ is distributed over a discrete constellation with one vector point $\mathbf{A}\mathbf{x}$ for each support $b$. The error probability can be upper-bounded by the union bound comprising the sum of all pair-vise error probabilities. The difference between two constellation vectors is the sum of products of the elements of $\mathbf{x}$ selected differently by the different supports and of the corresponding columns of $\mathbf{A}$.

Next we derive the pair-vise probability averaged over $\mathbf{A}$ of the nearest neighbor. Define the nearest neighbor as the error event of confusing one true support element $k$ with a false support element $m$. The change in $\mathbf{y}$ will be $\Delta \mathbf{y} = x_m \mathbf{a}_m - x_k \mathbf{a}_k$ where $\mathbf{a}_n$ is the $n$'th column of $\mathbf{A}$. Thus $\Delta \mathbf{y}$ is a weighted sum of two Gaussian vectors $\mathbf{a}_n$ and has the conditional PDF averaged over $\mathbf{A}$:

$$P(\Delta \mathbf{y}|\mathbf{x}) = N^p(0, x_m^2 + x_k^2) \qquad (19)$$

which denotes here a length-p vector with i.i.d. complex Gaussian elements each with variance of $2\sigma^2 = |x_m|^2 + |x_k|^2$. Note that σ depends on **x**.

The following analysis benefits from considering **y** as a length $2p$ vector with real i.i.d. elements of variance $\sigma^2$ each. The PDF of the normalized length of $\Delta\mathbf{y}$, $d = |\Delta\mathbf{y}|/\sigma$, is the Chi distribution with $2p$ degrees of freedom independent of σ:

$$P(d) = \frac{1}{\Gamma(p)} 2^{1-p} d^{2p-1} e^{-d^2/2} \quad (20)$$

The probability of error conditioned on $d$ and on σ is the probability that the component $z_d$ of the Gaussian noise **z** in the direction of $\Delta\mathbf{y}$ will exceed $0.5d\sigma$. The noise is isotropic, thus the variance of its projection $z_d$ in the direction of $\Delta\mathbf{y}$ is $0.5q/SNR$ and the conditional probability of error is

$$P(er|d,\sigma) = P(z_d \geq 0.5 \mathrm{d}\sigma)$$
$$= \frac{1}{2} erfc(\frac{0.5\mathrm{d}\sigma}{\sqrt{0.5q/SNR} * \sqrt{2}})$$
$$= \frac{1}{2} erfc(\frac{0.5\mathrm{d}\sigma\sqrt{SNR}}{\sqrt{q}}) \quad (21)$$

The probability of error is the average of the last equation:

$$P(er) = \iint_{d\ \sigma} P(er|d,\sigma)P(d)P(\sigma)\mathrm{d}\sigma\mathrm{d}d \quad (22)$$

where σ is Chi distributed with 4 degrees of freedom, each with variance of 0.25:

$$P(\sigma) = 2\frac{1}{\Gamma(2)} 2^{-1}(2\sigma)^3 e^{-(2\sigma)^2/2} = 8\sigma^3 e^{-2\sigma^2} \quad (23)$$

Eq. (22) is evaluated by numerical integration and presented in figure 3 below, denoted 'Analytic, single nearest neighbor'.

The number of possible nearest neighbors is $q(n-q)$ thus an approximate union bound discarding all neighbors except all the nearest ones is

$$P_{ub} = q(n-q)P(er) \quad (24)$$

An accurate union bound can be derived in a straightforward manner by the same technique; however the simulation below shows that this is not necessary, at least for the parameters used in the simulation, since most of the errors are of the nearest neighbor type.

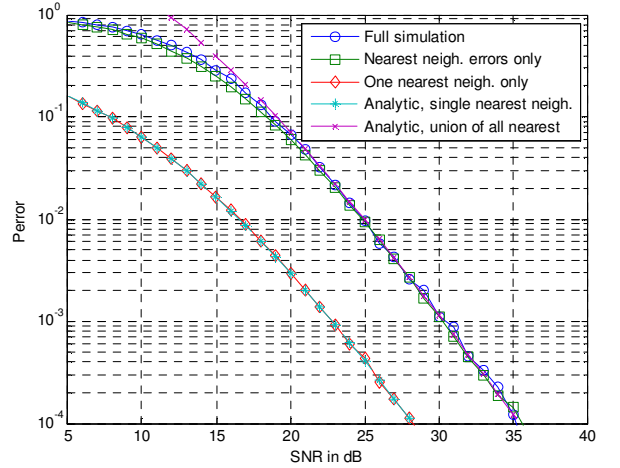

Figure 3. Probability of a support detection error. Solid lines are simulation results, dashed lines are eq. (22), (24). *SNR* is per element of **y**. $n$=10, $q$=6, $p$=3.

The support detection was simulated assuming correct detection of the coded signal *c*. The detection was a maximal likelihood search for a signal point **Ax** with smallest Euclidian distance to the received signal **y** which is practical for moderate *q* and *n* while for large values an iterative decoder might be developed along the lines of [22]. The results are denoted 'Full simulation' in figure 3. Since the support itself is not coded its detection is error prone, however at high enough SNR the probability of error is very low in our example which operates at half the Landau sampling rate. Additional two genie-aided lines were generated. On the first one the receiver had to decide only between the true signal and all its nearest neighbors, this is denoted 'Nearest neighbors errors only'. The performance is almost identical to the full simulation showing that most errors are of the nearest neighbor type, thus misdetecting only one element of the support. Another line denoted 'One nearest neighbor only' was generated while choosing merely between the true signal and one of the nearest neighbors, this lowered the error probability as expected. A line 'Analytic, single nearest neighbor' according to the analytic probability of error (22) is plotted, it is practically identical to the line simulated with the single neighbor, verifying(22). Another line is eq. (24), denoted 'Analytic, union of all nearest', it is very close to the true performance at SNRs high enough verifying the approximate union bound.

The support recovery would benefit from non-Gaussian signals guaranteed to exceed some absolute value similar to some signals examined in [2] since this would improve the PDF of σ in (23) reducing the probability of support detection error. Indeed the tradeoff between the support estimation performance and the information rate $R_c$ is derived in [20, theorem 1] which shows that the support (side information in [20]) detection can be improved by imposing constraints on the transmitted signals at the cost of some decrease in $R_c$.

## 5 THE ASYMPTOTIC HIGH-DIMENSIONAL CASE

Let us consider the asymptotic case with $p$, $q$ and $n$ approaching infinity while keeping their ratios with respect to $n$ constant:

$$p_R = \frac{p}{n}, \quad q_R = \frac{q}{n}$$

For large $n$ and $q$ the known asymptotic behavior of the entropy of the support (6) as obtained by the Stirling's approximation is

$$H(b) \approx n\,\mathrm{H}(q_R) \tag{25}$$

Where $\mathrm{H}(q_R) = -q_R \log(q_R) - (1-q_R)\log(1-q_R)$ denotes the binary entropy function of $q_R$. As shown above,

$$H(b) + R_c \leq I(\mathbf{x};\mathbf{y}\,|\,\mathbf{A}) \tag{26}$$

is a necessary condition for a joint support and signal detection. When the support is known to the transmitter this condition is also sufficient for reliable detection as shown in section 3 above. Thus the results on the required SNR derived in this section apply to the case of support known to the transmitter. When the support is not known at the transmitter it provides the necessary condition but not the support detection error probability.

As explained above and as seen in figure 1, when working significantly below the Landau rate, such as $p < 0.5q$, the upper bound on the mutual information $I(\mathbf{x};\mathbf{y}\,|\,\mathbf{A})$ of the channel (11) and the lower bound $I_{MIMO}$ (10) are separated only by a small gap, therefore both are good practical approximations of the actual mutual information. Next we shall show using the results of [14] and [15] that this holds also for large $p$ and $q$ for a wide range of matrices $\mathbf{A}$ including the two types we used in this work. The asymptotic high dimensional analysis of $\mathbf{A}$ with i.i.d. elements, including the Gaussian distribution used here is presented in [14]; for simplicity we use here the large *SNR* results. The gap between $I_{MIMO}$ available from [14], to the upper bound (11) is $p \cdot C_{gap}^G$ where

$$C_{gap}^G = \log e + (\beta - 1)\log\frac{(\beta-1)}{\beta} + o(1) \tag{27}$$

and $\quad \beta = \dfrac{q_R}{p_R} > 1$

See the appendix for a more detailed reference to [14]. Note that $C_{gap}^G$ is independent of SNR, thus it may be discarded when approximating at large SNR. Results for general SNR are available in [14], those would yield a more complex expression of $C_{gap}^G$. Applying (27) to our small dimensional example in fig. 1 of $q=3$, $p=3$, $\beta=6/3=2$ yields $p \cdot C_{gap}^G$ of 1.3281 bits, very similar to the gap seen on figure 1 for *SNR* > 15 dB. This indicates that the high dimensional result is a good approximation already with small number of dimensions as in other sections in [14] and that (11) approximates $I(\mathbf{x};\mathbf{y}\,|\,\mathbf{A})$ for large $p$ about as well as the curves in Fig. 1 generated for $p=3$.

The high dimensional case with $\mathbf{A}$ being a submatrix of the discrete Fourier matrix with the rows and columns chosen randomly and uniformly is treated in [15] as shown in the appendix; again we use the large SNR results while the general SNR results are available in [15]. The gap between $I_{MIMO}$ in [15] to the upper bound (11) is $p \cdot C_{gap}^F$,

$$\begin{aligned}C_{gap}^F =\ & \frac{-p_R \log_2 p_R - (1-p_R)\log_2(1-p_R)}{p_R} \\ & + \log_2\frac{p_R}{q_R - p_R} - \frac{q_R}{p_R}\log_2\frac{q_R}{q_R - p_R} + \log_2(q_R)\end{aligned} \tag{28}$$

Again $C_{gap}^F$ is independent of SNR. Interestingly it vanishes at $q_R=1$, see the appendix. Applied to our example in figure 1 of $q_R=0.6$, $p_R=0.3$ it yields 0.4 bits per channel use as compared to about 0.6 bits in figure 1 demonstrating again that the high-dimensional result is a usable approximation also in the low dimensional case.

Expressions on $I(\mathbf{x};\mathbf{y}\,|\,\mathbf{A},b)$ denoted $I_2$ in [3] for additional sensing matrices are derived in [3].

Thus [14] and [15] with (11) provide the large $p$, $q$ asymptotic form:

$$I(\mathbf{x};\mathbf{y}|\mathbf{A}) \geq I_{MIMO} \approx p[\log(1+SNR) - C_{Gap}] \quad (29)$$

with $C_{Gap}$ moderate and not growing with the SNR. Combining this lower bound on $I(\mathbf{x};\mathbf{y}|\mathbf{A})$ with the approximation (25) and with (26) yields then the following approximate upper bound on the required SNR:

$$n\,\mathrm{H}(q_R) + R_c \approx p\,[\log(1+SNR) - C_{Gap}]$$

$$SNR \approx 2^{\frac{\mathrm{H}(q_R) + R_{cp} + C_{Gap}}{p_R}} - 1 \quad (30)$$

with $R_{cp} = R_c / p$ being the signal information rate in bits per sample. Deleting $C_{Gap}$ from this equation and from all the similar equations below provides, by (11), upper bounds on mutual information and lower bounds on the required SNR. We use the $\approx$ sign to denote those approximate asymptotic bounds. Clearly very high SNR is needed at very low $p_R$ relevant to very sparse scenarios.

Approximating $\mathrm{H}(q_R) \approx -q_R \log(q_R)$ which is valid at very sparse scenarios at which $q_R \ll 1$, yields

$$SNR \approx \left(\frac{1}{q_R}\right)^{\frac{q_R}{p_R}} 2^{C_{gap} + R_{cp}} + 1 \quad (31)$$

So finite but very high SNR is required for reliable operation in very sparse scenarios with low sampling rate even at the minimal signal information rate $R_c$=0 and reducing the sampling rate $p_r$ inflates the required SNR further.

Sampling rate below the Landau rate for coded signals at moderate SNR is still applicable to some very sparse practical communication systems in which the entropy of the frequency support is limited significantly below (6). The trivial example is of course a known spectral occupancy with a zero entropy which can occur in a practical scenario. More interesting application is when the frequency spectra is assigned in finite number of subbands as common in radio communication systems.

Let the total available bandwidth $W$ be divided into $K$ subbands of bandwidth $W/K$ each. The signal occupies $q_R K$ subbands selected uniformly for each time interval $T \gg K/W$. Then we have in each interval $WT$ Nyquist rate samples and $q_R WT$ Landau rate samples. If sampling at a slower rate of $p_R W$ the mutual information per interval with Gaussian signaling will be by (29) close to $p_R WT \left[\log(1+SNR) - C_{gap}\right]$ while the entropy of the support for large $q_R K$ will be

$$H(b) = \log\binom{K}{q_R K} \approx K\,\mathrm{H}(q_R) \quad (32)$$

which does not grow with $W$ or with $T$. Reliable detection will be possible if the mutual information exceeds the sum of the signal information rate and the entropy of the support. Approximating the mutual information by (29) yields an approximation on the possible sampling rate valid for large $q_R K$:

$$K\,\mathrm{H}(q_R) + R_c \approx p_R WT \left[\log(1+SNR) - C_{gap}\right]$$

$$SNR \approx 2^{\frac{K}{WT}\frac{\mathrm{H}(q_R)}{p_R}} 2^{C_{gap} + R_{cp}} - 1 \quad (33)$$

The first fraction in the equation is the inverse of the number of frequencies resolvable in each subband which reduces the exponent, so reliable detection below the Landau rate is possible in many practical systems at moderate SNR if the number of subbands is limited.

Similar exploitation of the reduction of entropy of the support to enable the support detection was presented in the sparse sampling literature, e.g. [6] and [7] where the frequency support is a finite union of bounded intervals and it is related to sparse sensing scenarios with multiple measurements [23].

## 6 CONCLUDING REMARKS

We examined sparse sampling and sparse sensing in the domain of coded signals and of sampling rates lower than the Landau rate. As expected the Landau condition [5] on sampling rates derived for continuous uncoded signals is not necessary when the signals are coded. Limiting the cardinality of the signal space by coding impacts directly the required sampling rate as compared to the Landau rate associated with the effective bandwidth only. The sum of the signal information rate and of the entropy of the support must not exceed the mutual information of the sparse channel; this requirement governs the tradeoff between the information rate, sparsity, SNR, a-priory uncertainty about the support at the detector and the sampling rate. We derived tight upper and lower bounds on the mutual information using known results over the MIMO channel and demonstrated that coded signals can be detected reliably with sampling rates lower than the Landau rate.

In the high-dimensional case the entropy of the support grows, thus necessitating high SNR similar qualitatively to the effect reported in [2] and [3]. We derived the bounds for this case which are different from those of [2] due to sampling below the Landau rate and due to exploiting the code to reduce the probability of support detection error. We showed that operation in the high-dimensional regime is still possible at moderate SNR in practical radio communication systems if the entropy of the support is reduced by assigning bandwidth by finite number of predefined frequency bands.

The results may be relevant also to the sparse sensing problem, e.g. identifying an object out of a finite cardinality class of objects using sparse sensing will require less measurements than producing an image of an arbitrary sparse object.

## APPENDIX: THE KNOWN ASYMPTOTIC RESULTS ON CAPACITY

### The Gaussian matrix A

The matrix of i.i.d. elements with asymptotically large $p$ and $q$ is treated in [14, equation (3.140)]. This includes the Gaussian distribution used here. The case of large SNR is treated in [14, two equations below (3.140)] and yields (27). Sampling below the Landau rate sets $\beta > 1$ which is used to select the proper case in [14].

### The Fourier submatrix A

As evident in [15], when the elements of the diagonal matrices **A** and **G** representing the frequency and time domain fading there are 0 or 1 the results are the capacity of a channel in the form of our (1) with **A** in (1) being a submatrix of a discrete Fourier matrix with randomly chosen rows and columns. Sampling below the Landau rate sets $q_R > p_R$ which is used to select the proper case in [15]. The general SNR is treated in [15, equation (18)]. To derive a simple intuitive expression we use the high SNR results in [15, eq. (45) to (50)]. The expressions relating our variables to those of [15] are

$$SNR = \gamma p, \quad u_0 = 1 - q_R, \quad v_0 = 1 - p_R, \quad u_0 < v_0$$

where $u_0, v_0, \gamma$ are variables from [15]. Solving [15, eq. (45)] using $P(G=1)=p_R$, $P(G=0)=1-p_R$ when evaluating expectation yields

$$\psi = \frac{p_R}{q_R - p_R}$$

Substituting to [15, eq. (47) and (48)] while accounting for the relation between our SNR and $\gamma$ in [15] and comparing to (11) yields (28). Interestingly the gap to (11) vanishes at $q_R=1$ which is explained by the orthogonality of the rows of the discrete Fourier matrix which render the elements of **y** independent.